\documentclass[12pt]{article}

\usepackage{amsmath,amssymb}
\usepackage{epsfig} 
\usepackage{xspace}  
\usepackage{refmerge}  

\def\m@th{\mathsurround=0pt }
\def\leftrightarrowfill{$\m@th \mathord\leftarrow \mkern-6mu
        \cleaders\hbox{$\mkern-2mu \mathord- \mkern-2mu$}\hfill
        \mkern-6mu \mathord\rightarrow$}

\def\overleftrightarrow#1{\vbox{\ialign{##\crcr
        \leftrightarrowfill\crcr\noalign{\kern-1pt\nointerlineskip}
        $\hfil\displaystyle{#1}\hfil$\crcr}}}

\newcommand{\newc}{\newcommand}
\newcommand{\renewc}{\renewcommand}
\newc{\tr}{{\rm Tr}}
\newc{\be}{\begin{equation}}
\newc{\ee}{\end{equation}}

\newc{\STr}{{\rm STr}}

\def\shat{\ifmmode \hat{s}\else $\hat{s}$\fi}
\def\gp2{{g'}^2}
\def\g2{g^2}
\def\g32{g_s^2}

\newc{\gsim}{\lower.7ex\hbox{$\;\stackrel{\textstyle>}{\sim}\;$}}
\newc{\lsim}{\lower.7ex\hbox{$\;\stackrel{\textstyle<}{\sim}\;$}}
\newc{\cf}{{\it cf.}\xspace}
\newc{\ie}{{\it i.e.}\xspace}
\newc{\eg}{{\it e.g.}\xspace}
\newc{\etal}{{\it et al.}\xspace}
\newc{\mev}{\hbox{\rm\:MeV}\xspace}
\newc{\gev}{\hbox{\rm\:GeV}\xspace}
\newc{\tev}{\hbox{\rm\:TeV}\xspace}
\newc{\xpb}{\hbox{\rm\: pb}\xspace}
\newc{\xfb}{\hbox{\rm\: fb}\xspace}

\newc{\G}{{\cal G}}
\newc{\h}{{\cal H}}
\newc{\D}{{\cal D}}
\newc{\E}{{\cal E}}

\newc{\dalphaHadMZ}{\ensuremath{\Delta\alpha_{\rm had}^{(5)}(M_Z^2)}\xspace}
\newc{\Dalphahad}{\dalphaHadMZ}
\newc{\as}{\ensuremath{\alpha_{\scriptscriptstyle S}}\xspace}
\newc{\asZ}{\ensuremath{\as(M_Z^2)}\xspace}
\newc{\seffsf}[1]{\sin\!^2\theta^{#1}_{{\rm eff}}}
\newc{\sinfeff}{\sin\!^2\theta^f_{{\rm eff}}}
\newc{\sinleff}{\seffsf{\ell}}
\newc{\CL}{\ensuremath{\rm CL}\xspace}

\newc\defaultFigureScale{0.65}

\newc{\mtop}{m_t}
\newc{\mbot}{m_b}
\newc{\mz}{M_Z}
\newc{\mw}{M_W}
\newc{\alphasmz}{\asZ}
\newc{\swsq}{\sin^2\theta_W}
\newc{\cwsq}{\cos^2\theta_W}
\newc{\tw}{\tan\theta_W}
\newc{\cw}{\cos\theta_W}
\newc{\sw}{\sin\theta_W}
\newc{\BR}{\hbox{\rm BR}}
\newc{\zbb}{Z\to b\bar}
\newc{\Gb}{\Gamma (Z\to b\bar b)}
\newc{\Gh}{\Gamma (Z\to \hbox{\rm hadrons})}
\newc{\sgn}{\mbox{sgn}}

\newc{\sep}[1]{#1}
\newcounter{mysubequation}[equation]
\renewc{\themysubequation}{\alph{mysubequation}}
\newc{\mytag}{\stepcounter{mysubequation}%
\tag{\theequation\protect\sep{\themysubequation}}}
\newc{\globallabel}[1]{\refstepcounter{equation}\label{#1}}

\newc{\msq}{m_{\tilde q}}
\newc{\msl}{m_{\tilde l}}
\newc{\MHc}{M_{H_C}}
\newc{\MAd}{M_\Sigma}

\def\beq{\begin{equation}}
\def\eeq{\end{equation}}
\def\bea{\begin{eqnarray}}
\def\eea{\end{eqnarray}}
\def\slashchar#1{\setbox0=\hbox{$#1$}           
   \dimen0=\wd0                                 
   \setbox1=\hbox{/} \dimen1=\wd1               
   \ifdim\dimen0>\dimen1                        
      \rlap{\hbox to \dimen0{\hfil/\hfil}}      
      #1                                        
   \else                                        
      \rlap{\hbox to \dimen1{\hfil$#1$\hfil}}   
      /                                         
   \fi}                                         %
\catcode`@=11
\long\def\@caption#1[#2]#3{\par\addcontentsline{\csname
  ext@#1\endcsname}{#1}{\protect\numberline{\csname
  the#1\endcsname}{\ignorespaces #2}}\begingroup
    \small
    \@parboxrestore
    \@makecaption{\csname fnum@#1\endcsname}{\ignorespaces #3}\par
  \endgroup}
\catcode`@=12

\begin{document}

\baselineskip=18pt

\setcounter{footnote}{0}
\setcounter{figure}{0}
\setcounter{table}{0}

\begin{titlepage}
\begin{flushright}
CERN-PH-TH/2009-058\\
\today \\
\end{flushright}
\vspace{.3in}

\begin{center}
{\Large \bf The Probable Fate of the Standard Model}

\vspace{0.5cm}

{\bf J. Ellis$^{a}$, J.R. Espinosa$^{a,b}$, G.F. Giudice$^{a}$, 
A. Hoecker$^{a}$} and 
{\bf A. Riotto$^{a,c}$}

\centerline{$^{a}${\it Physics Department, CERN, CH--1211 Geneva 23, 
Switzerland}}
\centerline{$^{b}${\it ICREA, Instituci\'o Catalana de Recerca i Estudis 
Avan\c{c}ats,}}
\centerline{\it at IFAE, Universitat Aut\`onoma de Barcelona, 08193 
Bellaterra, 
Barcelona, Spain}
\centerline{$^{c}${\it INFN, Sezione di Padova, Via Marzolo 
8, I-35131 Padua, Italy}}

\end{center}
\vspace{.8cm}

\begin{abstract}
\medskip
\noindent
Extrapolating the Standard Model to high scales using the renormalisation
group, three possibilities arise, depending on the mass of the Higgs
boson: if the Higgs mass is large enough the Higgs self-coupling may blow
up, entailing some new non-perturbative dynamics; if the Higgs mass is
small the effective potential of the Standard Model may reveal an
instability; or the Standard Model may survive all the way to the Planck
scale for an intermediate range of Higgs masses. 
This latter case does not necessarily require stability at all times, but 
includes the possibility of a metastable vacuum which has not yet decayed. 
We evaluate the relative likelihoods of these possibilities, on the basis of 
a global fit to the Standard Model made using the Gfitter package. This uses the
information about the Higgs mass available directly from Higgs searches at
LEP and now the Tevatron, and indirectly from precision electroweak data.
We find that the `blow-up' scenario is disfavoured at the 99\% confidence
level (96\% without the Tevatron exclusion), whereas the `survival' and 
possible `metastable' scenarios remain plausible. A future measurement 
of the mass of the Higgs boson could reveal the fate of the Standard Model.
\end{abstract}

\bigskip
\bigskip

\end{titlepage}


\section{Introduction}

The success of the Standard Model (SM) offers very few experimental clues
how it may break down, and at what scale. One clue is provided by the
discovery of neutrino masses, which suggest the appearance of new physics
at a mass scale of a TeV or more, probably at least $10^{10}\gev$ in the
simplest versions of seesaw models. Another clue might be offered by the
measurement of the anomalous magnetic moment of the muon, if one could be
sure of the value within the SM. However, this requires input from data on
low-energy $e^+ e^-$ annihilation and/or $\tau$ decay into hadrons about
which there is, unfortunately, as yet no consensus. The existence of dark
matter could be another clue to physics beyond the SM, assuming it does
not have some astrophysical origin such as primordial black holes. The
baryon asymmetry of the Universe can also be explained only by physics
beyond the SM, which could appear anywhere between the electroweak and
inflation scales.

In view of this paucity of experimental hints about possible physics
beyond the SM, any new indications would be most welcome. We discuss in
this paper the one important hint about the possible scale of new physics
that may (soon) be provided by the Higgs sector of the SM. There are, of
course, plenty of theoretical arguments why the Higgs sector of the SM is
inadequate, many of them related to the apparently unnatural fine-tuning
of its parameters, but we have in mind a more direct empirical argument
based on the available experimental information about the Higgs sector.

\begin{figure}[t]
\begin{center}
  \epsfig{file=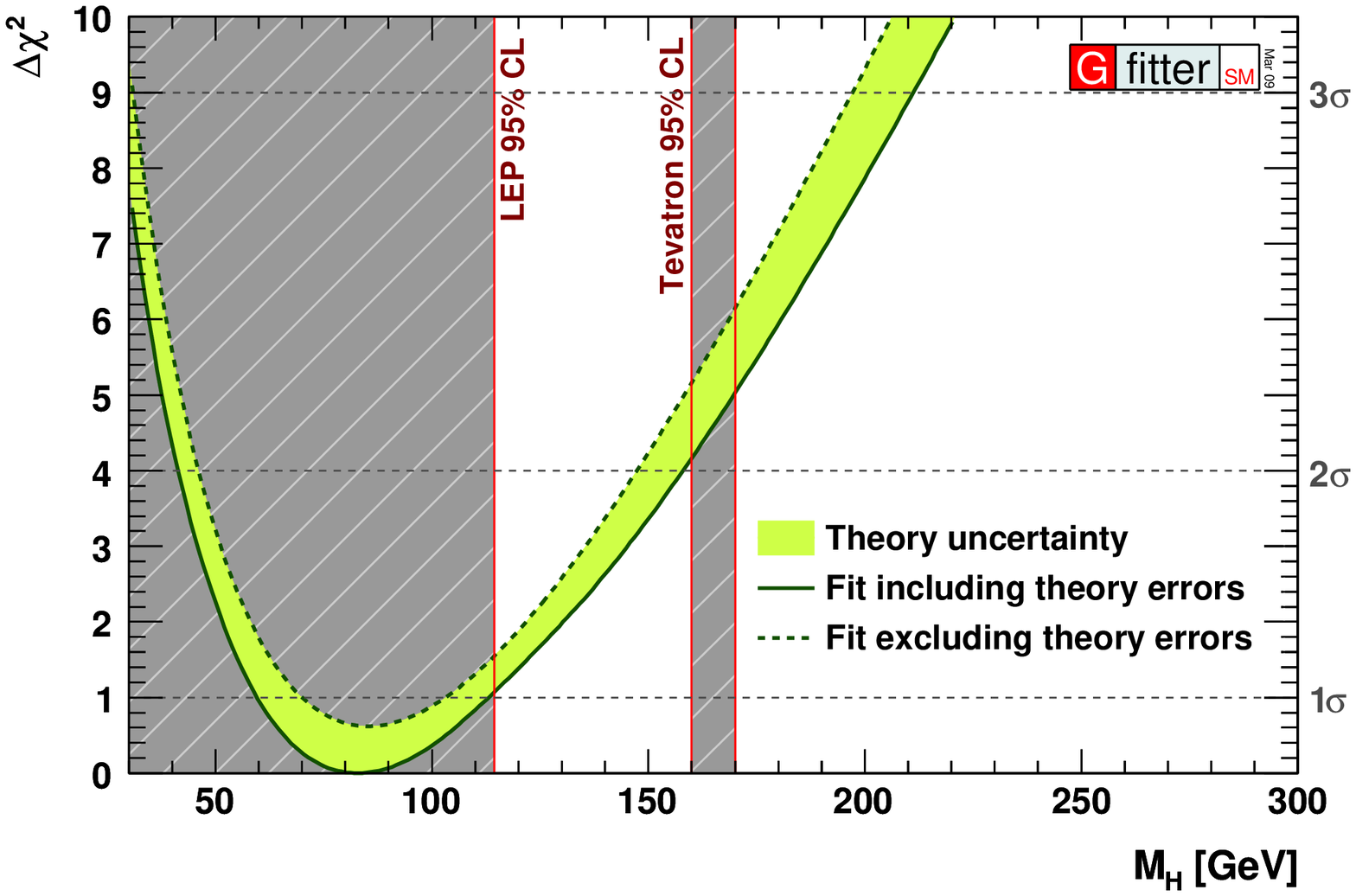, scale=0.405}
  \epsfig{file=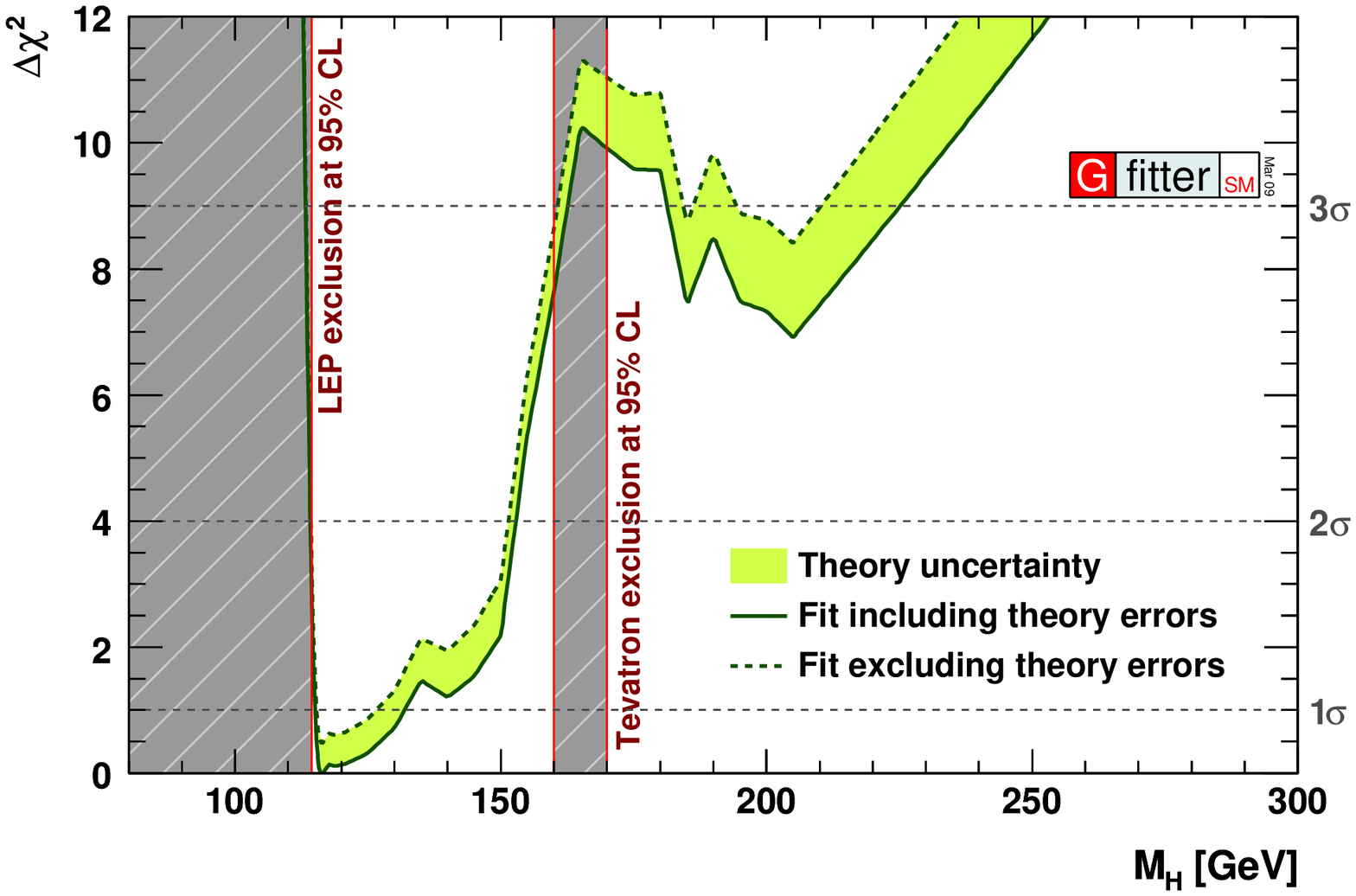, scale=0.405}
\end{center}
  \vspace{-0.5cm}
\caption{\it  Dependence on $M_H$ of the $\Delta \chi^2$ function obtained 
         from the global fit of the SM parameters to precision 
         electroweak data~\cite{gfitter}, excluding (left) or including (right) the 
         results from direct searches at LEP and the Tevatron.
\label{fig:basis}}
\end{figure}

The most direct information comes from experimental searches for the SM
Higgs boson, first at LEP and more recently at the Tevatron. These exclude
a Higgs mass $M_H < 114.4\gev$~\cite{Higgs-LEP} and between 160 and
170\gev~\cite{Higgs-Tev} at the 95\% confidence level (CL), and also provide
contributions to the overall SM likelihood function for other
values of the Higgs mass. Another contribution to the Higgs likelihood
function comes from a global fit to electroweak precision data within the SM,
which favours $M_H < 158\gev$~\cite{gfitter} (95\% CL, not including the 
direct Higgs searches). Figure~\ref{fig:basis} shows 
the $\Delta\chi^2$ function obtained from the global fit without (left
hand plot) and with (right) the information from the direct Higgs searches 
at LEP and the Tevatron.

\begin{figure}[t]
\begin{center}
  \epsfig{file=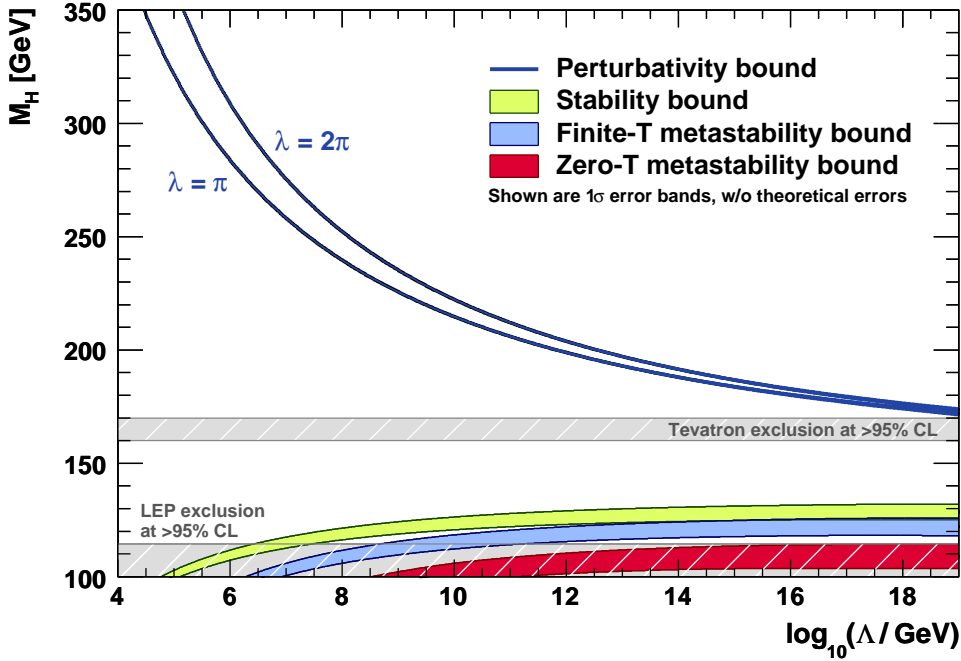, scale=\defaultFigureScale}
\end{center}
  \vspace{-0.5cm}
\caption{\it The scale 
         $\Lambda$ at which the two-loop RGEs drive the quartic SM Higgs 
         coupling non-perturbative, and the scale $\Lambda$ at which the RGEs 
         create an instability in the electroweak vacuum ($\lambda < 0$). 
         The width of the bands indicates the errors 
         induced by the uncertainties in $\mtop$ and $\as$ (added quadratically).         
         The perturbativity upper bound (sometimes referred to as 
         `triviality' bound) is given for $\lambda = \pi$ 
         (lower bold line [blue]) and $\lambda =2\pi$ 
         (upper bold line [blue]). Their difference indicates the size of the
         theoretical uncertainty in this bound. The absolute 
         vacuum stability bound is displayed by the light shaded [green] band,
         while the less restrictive finite-temperature and zero-temperature 
         metastability bounds are medium [blue] and dark shaded [red], respectively.
         The theoretical uncertainties in these bounds have been ignored in 
         the plot, but are shown in Fig.~\protect\ref{fig:bounds_zoom} (right panel). 
         The grey hatched areas indicate the LEP~\protect\cite{Higgs-LEP} and 
         Tevatron~\protect\cite{Higgs-Tev} exclusion domains.
\label{fig:bounds}}
\end{figure}

\begin{figure}[t]
\begin{center}
  \epsfig{file=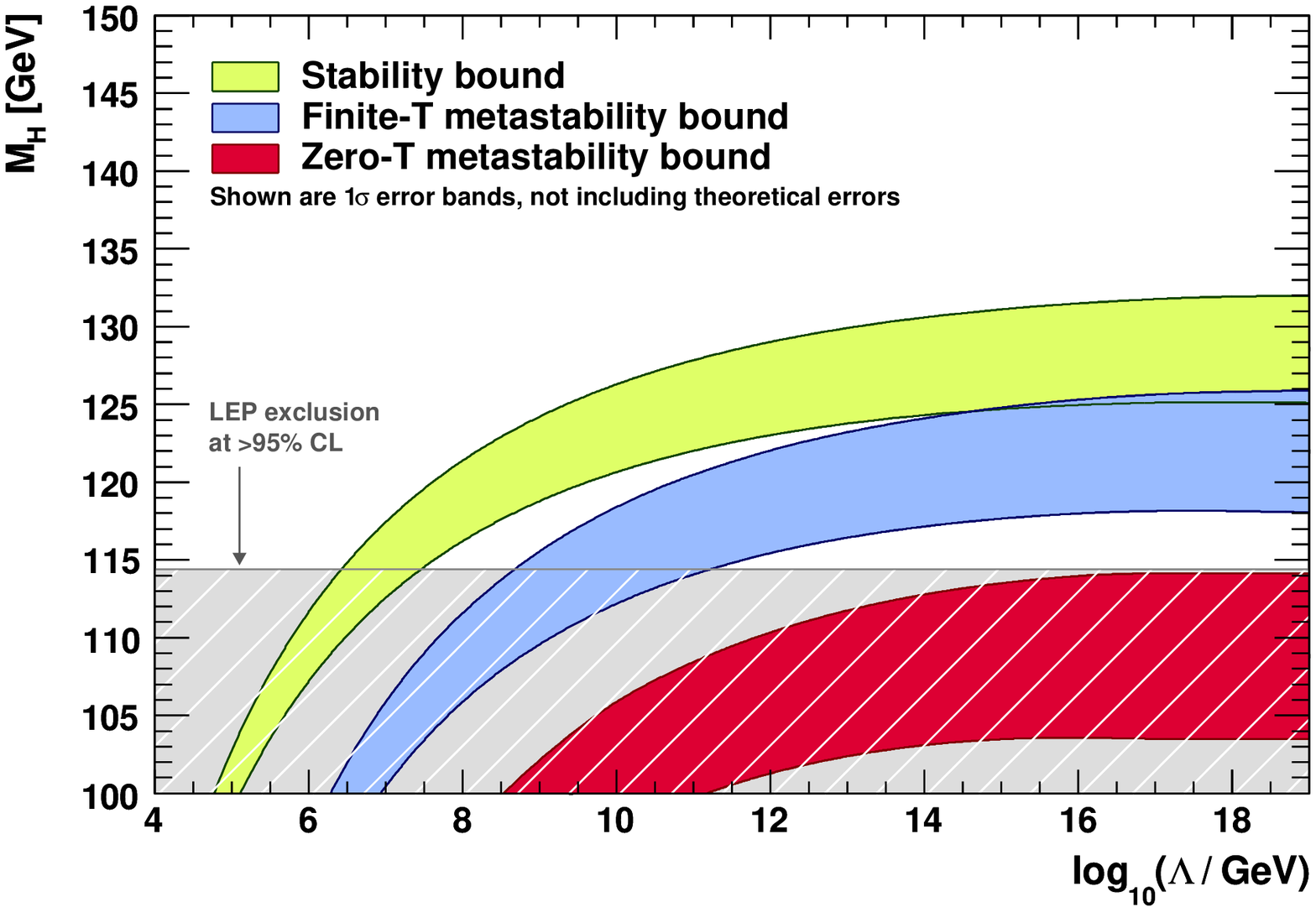, scale=0.405}
  \epsfig{file=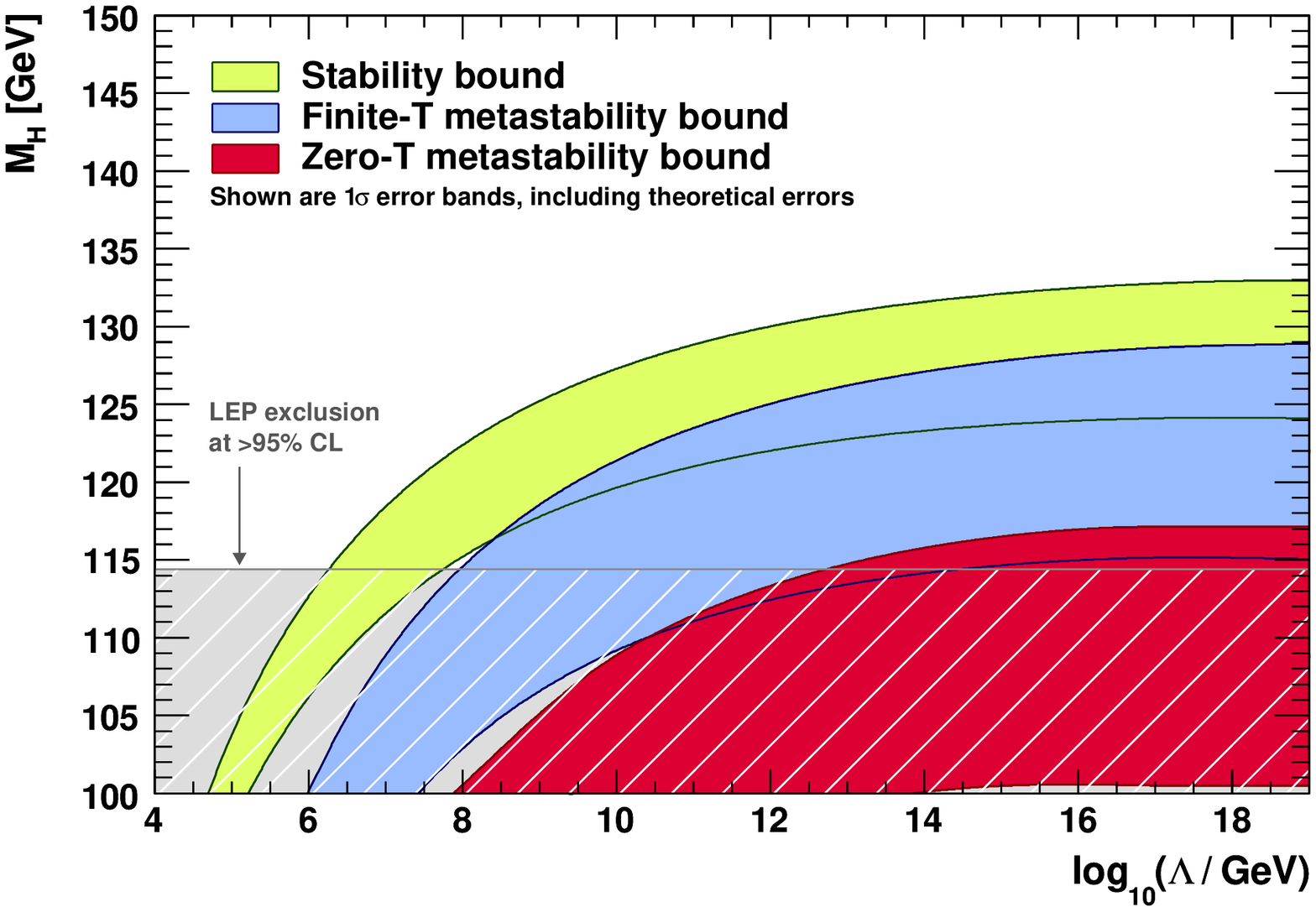, scale=0.405}
\end{center}
  \vspace{-0.5cm}
\caption{\it Lower bounds on the Higgs mass due to 
absolute vacuum stability (light shaded [green]), finite-temperature 
         (medium shaded [blue]) and zero-temperature metastability (dark 
         shaded [red]), as functions of the cut-off scale 
         $\Lambda$. The bands indicate the errors induced by the uncertainties in 
         $\mtop$ and $\as$ (added quadratically). The left plot is thus identical 
         to Fig.~\ref{fig:bounds}, but with a zoomed ordinate. The right plot includes 
         theoretical uncertainties, which treated as an offset, \ie, they are not quadratically 
         added to the other errors (\cf Sec~\ref{sec:combinedLikelihoodAnalysis}).
         At $\Lambda=M_P$, the bounds correspond to Eqs.~(\ref{eq:stability}), 
         (\ref{eq:thermmetastab}) and (\ref{eq:metastab}), respectively. 
\label{fig:bounds_zoom}}
\end{figure}

It is well known that the Higgs sector of the SM must steer a narrow course
between two problematic situations if it is to survive up to the reduced
Planck scale $M_P \sim 2\times 10^{18}\gev$, by which some new physics
associated with quantum gravity must surely appear
\cite{con,Str,CEQ,Hambye,KoldMu}.  If $M_H$ is large enough, the
renormalisation-group equations (RGEs) of the SM drive the Higgs self-coupling
into the non-perturbative regime at some scale $\Lambda < M_P$, entailing
either new non-perturbative physics at a scale $\sim \Lambda$, or new physics
at some scale $< \Lambda$ that prevents the Higgs self-coupling from blowing
up. This is shown as the upper pair of bold [blue] lines in
Fig.~\ref{fig:bounds}. On the other hand, if $M_H$ is small enough, the RGEs
drive the Higgs self-coupling to a negative value at some Higgs field value
$\Lambda < M_P$, in which case the electroweak vacuum is only a local minimum
and there is a new, deep and potentially dangerous minimum at scales
$>\Lambda$. The electroweak vacuum can potentially become unstable against
collapse (either because of zero-temperature (quantum) or thermal tunneling
during the evolution of the universe) into that deeper new vacuum with Higgs
vacuum expectation value $> \Lambda$, unless there is new physics at some
scale $< \Lambda$ that prevents the appearance of that vacuum. This is shown,
with its uncertainties, as the light shaded [green] bands in
Figs.~\ref{fig:bounds} and \ref{fig:bounds_zoom}.  Below this stability bound,
there is a region we dub the `metastability' region where the electroweak
vacuum has a lifetime longer than the age of the Universe for decay via either
zero-temperature quantum fluctuations (region above the dark shaded [red]
bands in these figures) or thermal fluctuations (region above the medium
shaded [blue] bands). Between the `blow-up' and `metastability' cases, there
is a range of intermediate values of $M_H$ for which the SM could survive up
to the Planck scale.


In this paper we update and complete previous calculations of these bounds on $M_H$,
and then make quantitative estimates of the relative likelihoods
of these `blow-up', `collapse', `metastable' and `survival' scenarios, on the basis of
a combined analysis of the information currently available about the
possible mass of the Higgs boson within the SM, including both
experimental and theoretical uncertainties. Our principal conclusion is
that the non-perturbative `blow-up' scenario is now disfavoured at the
99.1\% CL after inclusion of the recent Tevatron exclusion of
a SM Higgs boson weighing between 160 and 170\gev~\cite{Higgs-Tev}, whereas this scenario
could only have been excluded at the 95.7\% CL if the Tevatron
information were not included. On the other hand, the Tevatron data, 
although able to narrow down the region of the `survival' scenario, have
no significant impact on the relative likelihoods of the `collapse', `metastable' and
`survival' scenarios, neither of which can be excluded at the present
time.

We also consider the prospects for gathering more information about the
fate of the SM in the near future. The Tevatron search for the SM Higgs
boson will extend its sensitivity to both higher and lower $M_H$, and then
the LHC will enter the game. It is anticipated that the LHC has the
sensitivity to extend the Tevatron exclusion down to $127\gev$ or less with
$1\:{\rm fb}^{-1}$ of well-understood data at 14\tev centre-of-mass energy~\cite{atlas-cscbook}. 
This would decrease the
relative likelihood of the `survival' scenario, but not sufficiently to
exclude it with any significance. On the other hand, discovery of a Higgs
boson weighing 120\gev or less would exclude the `survival' scenario with
high significance, implying the presence of a potential instability of 
the SM at some scale $\Lambda < 10^{10}\gev$, below the scale for new physics
that is suggested by simple seesaw models of neutrino masses.\footnote
{
   If the seesaw scale $M$ were higher than $\sim 10^{12}\gev$ the 
   stability and perturbativity bounds would get significantly more 
   stringent above $M$~\cite{seesawbounds}.
}

\section{Calculation of the SM Higgs Mass Bounds}
\label{sec:sm_bounds}

The SM effective potential for the real Higgs field $h$ can be written in
the 't Hooft-Landau gauge and the $\overline{\rm MS}$ renormalisation
scheme as $V=V_0+V_1$, where the tree-level $V_0$ and one-loop $V_1$
potentials are given by
\begin{eqnarray}
V_0&=&-\frac{1}{2} m(\mu)^2 
h^2(\mu)+\frac{1}{4}\lambda(\mu)h^4(\mu)\,,\nonumber\\
V_1&=&\sum_i \frac{n_i}{64\pi^2}M_i^4(h)\left[\log\frac{M_i^2(h)}{\mu^2}-
C_i\right]\,.
\end{eqnarray}
The sum is over all SM particles acquiring a Higgs-dependent mass $M_i(h)$
and having $n_i$ degrees of freedom (taken negative for fermions). The
coefficients $C_i$ are 5/6 (3/2) for gauge bosons (scalars and fermions), 
see
Ref.~\cite{CEQ} for more details.

Following Ref.~\cite{Espinosa:2007qp}, we work with the Higgs one-loop
effective potential improved by two-loop RGEs that resum contributions up
to next-to-leading logarithms~\cite{bando}.  The scale independence of the
effective potential $V$ allows us to fix the renormalisation scale $\mu$
at will for different values of the field \cite{bando,Casas:1994us}.  
Since our considerations refer to large field values, for our purposes it
is appropriate to choose the renormalisation scale to be the value of the
Higgs field, and to neglect the bilinear term. The SM Higgs potential is
therefore well approximated by
\beq
V(h)=\frac{\lambda (h)}{4}h^4\ ,
\label{V}
\eeq
where the running quartic coupling absorbs the large logs and includes in
its definition a one-loop finite non-logarithmic piece (see
Ref.~\cite{CEQ} for more details). The quartic Higgs coupling $\lambda$
and the top-quark Yukawa coupling $h_t$ that enter the RG evolution are
related to the physical Higgs and top pole masses through well-known
expressions that can be found, \eg, in the Appendix of
Ref.~\cite{Espinosa:2007qp}.

\begin{figure}[!Tt]
\begin{center}
  \epsfig{file=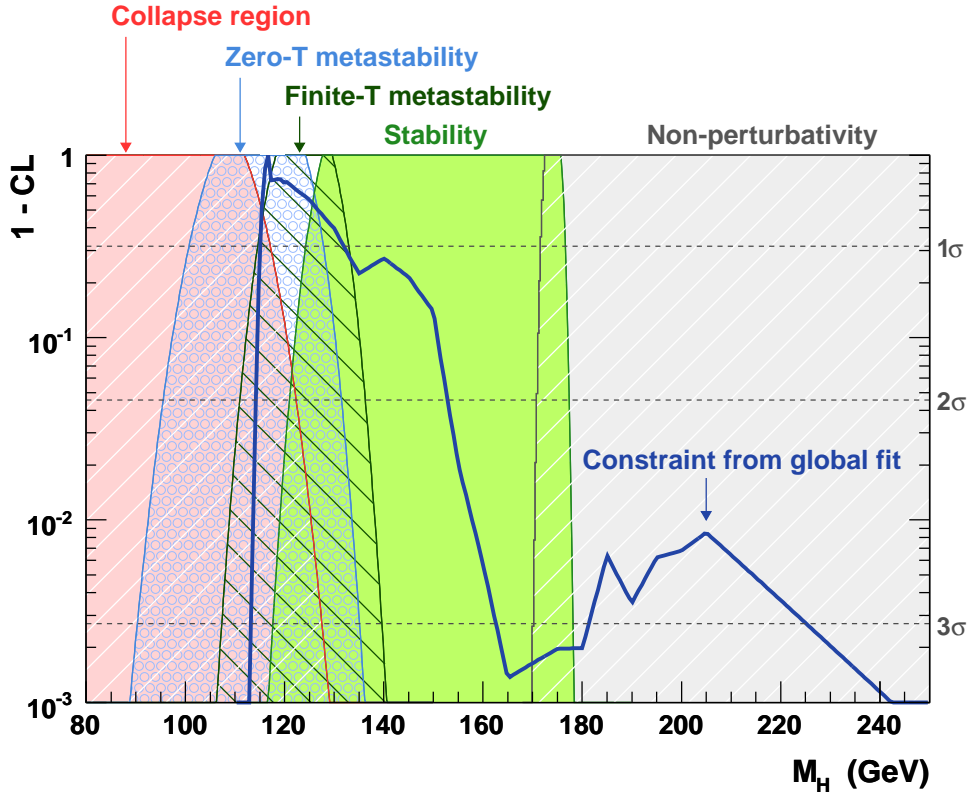, scale=\defaultFigureScale}
\end{center}
  \vspace{-0.5cm}
\caption{\it The levels of 1\,$-$\,CL versus $M_H$ for the different scenarios defined by
         the ultraviolet behaviour of the Higgs potential. The regions are (from left to right):
         the `collapse region' (light [red] shaded/hatched) corresponding to $M_H$
         violating the metastability bound~(\ref{eq:metastab}) and thus vulnerable to 
         quantum tunneling of the electroweak vacuum in a time shorter than the age 
         of the Universe; the `zero-temperature metastability' region
         ([blue] dotted) corresponding to values of $M_H$ between the bounds~(\ref{eq:metastab})
         and (\ref{eq:stability}), where quantum tunneling is acceptably slow; 
         the `finite-temperature metastability' region (dark [green] hatched), 
         defined by the lower bound~(\ref{eq:thermmetastab}), where the local
         SM minimum is stable against thermal fluctuations up to temperatures equal to $M_P$;
         the `stability' region
         (darker [green] shaded) delimited by the bounds~(\ref{eq:stability}) and (\ref{eq:blow-up});
         and finally the `non-perturbativity' region (light [grey] shaded/hatched), 
         bound by Eq.~(\ref{eq:blow-up}), where the Higgs self-coupling becomes 
         non-perturbative at some scale smaller than $M_P$.
         The slopes of the `pyramids' representing the boundaries of the different regions 
         reflect the uncertainties in 
         $\mtop$ and $\asZ$ which lead, together with the theoretical errors affecting the
         bounds, to apparent overlaps between the regions. Also shown is the 1\,$-$\,CL
         function for the combination of current constraints
         on $M_H$ equivalent to the right plot of Fig.~\ref{fig:basis} (bold solid [blue]
         line). }
\label{fig:MH_Lambda18}
\end{figure}

Following Ref. \cite{Hambye}, to compute the non-perturbativity bound we 
define two different conditions for the
scale $\Lambda$ at which we cut off the running: $\lambda_c(\Lambda)=\pi$
and $2\pi$. The first choice, $\lambda_c(\Lambda)=\pi$, corresponds to a
two-loop correction to the one-loop beta function $\beta_\lambda$ of the
Higgs quartic coupling of about 25\%, and the perturbative expansion is
still meaningful. The second choice, $\lambda_c(\Lambda)=2\pi$,
corresponds to a two-loop correction to $\beta_\lambda$ of about 50\%. The
bold [blue] upper lines in Fig.~\ref{fig:bounds} show the scale
$\Lambda$ at which the two-loop RGEs drive the quartic SM Higgs coupling
to the values $\lambda = \pi$ and $2 \pi$.  The (small) width of the lines
represents the the errors induced by the uncertainties in $\mtop$ and $\as$
(see below). Values above these lines
define the `blow-up' region where, for a given value of the Higgs mass,
either there is a scale $\Lambda$ at which some new non-perturbative
dynamics must appear, or there is some scale $<\Lambda$ where new physics
appears to avert the blow-up of the Higgs quartic coupling.  If we
require that this blow-up scale $\Lambda$ be larger than the reduced
Planck scale $M_P$, so that the SM remains in the perturbative regime, we find
\beq
\label{eq:blow-up}
M_H < M_H^c 
      + 0.7 \gev \left( \frac{\mtop -173.1 \gev}{1.3\gev} \right) 
      - 0.4\gev  \left( \frac{\asZ-0.1193}{0.0028}\right)
      \pm 1\gev
\eeq
with $M_H^c=175\gev$ (173\gev) for $\lambda(M_P)=2\pi$ ($\pi$).
We display explicitly the dependencies on the two most important SM parameters,
$\mtop$ and $\asZ$, normalising their effects in units of one standard
deviation from their experimental central values, for which we use $\mtop= 173.1 \gev
\pm 1.3\gev$~\cite{mtop} and $\asZ=0.1193 \pm 0.0028$~\cite{gfitter} throughout
this paper. The third (theoretical) error estimates the uncertainties from 
higher-order corrections in the running and matching of $\lambda$.
Figure~\ref{fig:MH_Lambda18} displays the $1 - \CL$ function at the
bound~(\ref{eq:blow-up}) as a narrow `pyramid' representing the
uncertain location of the boundary between the stable and
non-perturbative regions. The slopes of its sides reflect
the uncertainties in $\mtop$ and $\asZ$, and its width at the top reflects 
the theoretical error, which includes the ambiguity in the choice for
$\lambda_c(\Lambda)$. The non-perturbative region at larger $M_H$ is 
shaded light [grey].

The requirement that the electroweak vacuum be the absolute minimum of the
potential, up to a Higgs field scale $\Lambda $, implies $\lambda( \mu )>0$
for any $\mu <\Lambda$. The light shaded [green] band in
Fig.~\ref{fig:bounds} shows the scale $\Lambda$ at which the RGEs would
create a second minimum deeper than the electroweak vacuum ($\lambda <
0$), leading to a possible instability of the SM potential. 
The width of the band is obtained by varying the top mass and the
value of $\asZ$ by their one-standard-deviation errors. 
Fig.~\ref{fig:bounds_zoom} shows zooms of the low-mass region of Fig.~\ref{fig:bounds}:
the left plot is identical apart from the change in scale, whereas the right plot includes
an estimate of the overall uncertainty due to higher-order corrections.
We estimate this uncertainty by adding in the numerical calculation the known,
but incomplete, higher-order corrections. The largest effect comes from 
the two-loop 
QCD correction to the top-quark pole mass, which amounts to to a shift in $M_H$ of about $1\gev$. 
Since this effect is much larger than the parametric estimate of higher-order corrections, 
we consider it as a conservative choice for the theoretical error.

Requiring that the 
SM cannot develop a minimum deeper than the electroweak vacuum for any scale
$\Lambda<M_P$, we obtain the following lower bound on the Higgs mass:
\beq
\label{eq:stability}
M_H > 128.6 \gev + 2.6 \gev \left( \frac{\mtop -173.1 \gev}{1.3\gev} \right) 
-2.2\gev  \left( \frac{\asZ-0.1193}{0.0028}\right) \pm 1\gev\ .
\eeq
The Planck-scale stability bound (\ref{eq:stability}) is also shown in 
Fig.~\ref{fig:MH_Lambda18}
as a (somewhat broader) $1 - \CL$ `pyramid'. 
Equations~(\ref{eq:blow-up}) and (\ref{eq:stability}) 
delimit between them the `survival' region (represented as the shaded [green] 
band in 
Fig.~\ref{fig:MH_Lambda18}), within which the SM can be safely extrapolated up to 
the Planck scale.

It should be noted that the `unstable' region is not necessarily
incompatible with our existence, as long as the electroweak
vacuum survives for a time longer than the age of the universe, before 
quantum tunneling. The total quantum tunneling probability $p$ throughout the 
period of the history of the Universe during which thermal fluctuations have been negligible
is given by $p=\max_{h<\Lambda} [V_U h^4 \exp \left(
-8\pi^2/3|\lambda (h)| \right)]$, where $V_U=\tau_U^4$ is the space-time
volume of the past light cone of the observable Universe, $\tau_U$ being
the lifetime of the Universe. Taking $\tau_U=13.7\pm 0.2$~Gyrs from the analysis of WMAP
data~\cite{Spergel} and $p<1$, one finds that the electroweak vacuum has a
sufficiently long lifetime as long as 
\beq
\label{eq:metastab}
   M_H > 108.9 \gev 
         +4.0\gev \left( \frac{\mtop -173.1 \gev}{1.3\gev} \right) 
         -3.5\gev \left( \frac{\asZ-0.1193}{0.0028}\right) \pm 3\gev\ .
\eeq
The error of 3\gev is estimated by combining uncertainties from
higher-order corrections and from the prefactor in $p$. This
constraint is the leftmost `pyramid' in Fig.~\ref{fig:MH_Lambda18}, and
the  `collapse' region at lower $M_H$ is light [pink] shaded and hatched. 
The `metastability' bound obtained considering zero-temperature fluctuations up to a
scale $\Lambda$ is plotted as a dark shaded [red] band in 
Figs.~\ref{fig:bounds} and \ref{fig:bounds_zoom},
where the theoretical error is included only in the right plot of the latter figure.
The present LEP lower bound already rules out most of the parameter region where 
the electroweak vacuum is dangerously unstable, although this hypothesis cannot yet 
be excluded. We find a p-value of 0.40 for it being compatible with the LEP result.

The `metastable' region above~(\ref{eq:metastab})
and below~(\ref{eq:stability}), although compatible with
observations, is rather critical from the cosmological point of view,
because the SM vacuum becomes sensitive to thermal 
or inflationary fluctuations present during the early stages of the
Universe~\cite{Espinosa:2007qp,ArkaniHamed:2008ym}.
The requirement of thermal metastability depends on the temperature
up to which standard Big Bang cosmology is assumed.
For instance, requiring
the local SM minimum to be stable against thermal fluctuations up to temperatures
as large as the Planck scale  translates
into the lower bound~\cite{Espinosa:2007qp}
\beq
\label{eq:thermmetastab}
  M_H > 122.0 \gev 
        +3.0\gev \left( \frac{\mtop -173.1 \gev}{1.3\gev} \right) 
        -2.3\gev \left( \frac{\asZ-0.1193}{0.0028}\right) \pm 3\gev\ .
\eeq
The $1- \CL$ function for this constraint is shown as the second `pyramid'
from the left in Fig.~\ref{fig:MH_Lambda18}. 
The `finite-temperature metastability' bound is computed as follows. For
fixed $M_H$ in the metastable region there is a calculable maximum
temperature that the electroweak minimum can stand without decaying by
thermal fluctuations. For temperatures above that maximum value the decay
will proceed through thermal nucleation of bubbles that excite the Higgs
field at a typical value $h_N$ in the instability region of the effective
potential. To prevent this from happening, the effective potential should
be modified at or below the scale $h_N$, which we therefore identify with
the cut-off scale $\Lambda$ corresponding to the metastability bound.
(Typically this $\Lambda$ is one order of magnitude larger than the
maximum temperature for thermal tunneling.) The resulting bound is plotted
as a medium shaded [blue] band in Figs.~\ref{fig:bounds} and
\ref{fig:bounds_zoom}, where the theoretical error is included only in the
right plot of the latter figure.

Also shown in Fig.~\ref{fig:MH_Lambda18} is the $1 - \CL$ function for the
combined current constraints on $M_H$~\cite{gfitter}, equivalent to the
right plot of Fig.~\ref{fig:basis}. Both catastrophic scenarios,
`collapse' and `non-perturbativity', are disfavoured by the current data,
though the former cannot be excluded yet. Numerical results combining the
theoretical bounds and available constraint on $M_H$ are given in the
following section.

\section{Combined Likelihood Analysis}
\label{sec:combinedLikelihoodAnalysis}

We now convolve the information obtained from the (absolute) stability lower
bound and the `blow-up' upper bound on $M_H$, as functions of $\Lambda$,
with a likelihood analysis of $M_H$ based on electroweak precision data
and the direct Higgs boson searches. The numerical analysis is performed
with the Gfitter package~\cite{gfitter}. The latest
experimental inputs have been used, including the new world-average top-mass
result from the Tevatron~\cite{mtop}, a preliminary $M_W$
average~\cite{gfitter} incorporating the most recent measurement from the D0
experiment~\cite{MW-D0}, and a new combination of upper limits on
production of the SM Higgs boson from the CDF and D0
experiments~\cite{Higgs-Tev}.

The global electroweak fit uses as inputs the masses and widths of the $Z$
and $W$ bosons, the $Z$ hadronic and leptonic decay ratios and
forward-backward asymmetries, measurements of the heavy quark masses, and
the running fine structure constant at the $Z$ mass. The strong coupling
constant $\asZ$ is determined by the fit. References to all experimental
results, their SM predictions and the theoretical uncertainties affecting
them are available in Ref.~\cite{gfitter}. We include results from the
direct Higgs boson searches at LEP~\cite{Higgs-LEP} as well as the
Tevatron~\cite{Higgs-Tev} in the fit. The statistical procedure follows
Ref.~\cite{gfitter}, where in particular a two-sided CL is
used\footnote
{
   The numerical differences in the interpretation of the results from 
   the direct 
   Higgs searches between a one-sided or two-sided CL,
   or a Bayesian treatment (direct use of the likelihood ratio $\ln\!Q$), 
   are minor for the present data~\cite{gfitter}.
} 
to estimate the deviation of the measured event yields from the SM hypothesis 
for given $M_H$. The floating variables in the global electroweak fit
are the coupling strength parameters $\Dalphahad$ and $\asZ$, the
$Z$-boson mass, the quark masses $\mtop$, $m_b$, $m_c$, the Higgs boson
mass $M_H$, and four parameters quantifying theoretical uncertainties in
the predictions of $M_W$, $\sinleff$, and in the form factors absorbing
the radiative corrections to the effective weak mixing angle and to the
effective vector and axial-vector couplings of the $Z$ boson to
fermion-antifermion pairs.

The constraints on $M_H$ from the global fit obtained by the
Gfitter Group are shown in Fig.~\ref{fig:basis}, without (left panel) and
with (right panel) inputs from the direct Higgs searches in the fit. 
The 95\% CL allowed range for the complete fit (\ie,
including the direct searches) is $[114,153]\gev$, and above this range
only the values between 180\gev and 224\gev are not yet excluded at 3
standard deviations or more.

We incorporate the constraints from the (absolute) vacuum stability and
perturbativity requirements numerically into Gfitter. In the case of the
vacuum stability bound, the dependence on the floating parameters $\asZ$
and $\mtop$ are parametrised linearly and included in the fit. Also
included is a universal theoretical error of $1\gev$ on the bound,
parametrising uncertainties from higher-order perturbative terms (\cf
Sec.~\ref{sec:sm_bounds}). This error, as all theoretical errors in
Gfitter, is treated as a fit parameter varying freely within the given
range, which corresponds to adding a likelihood term to the fit function
that is finite and uniform within this range and zero outside.
For the perturbativity bound, we use the more conservative choice
$\lambda_c(\Lambda)=2\pi$ (\cf Sec.~\ref{sec:sm_bounds}).  
Other theoretical errors are neglected.

\begin{figure}[p]
\begin{center}
  \epsfig{file=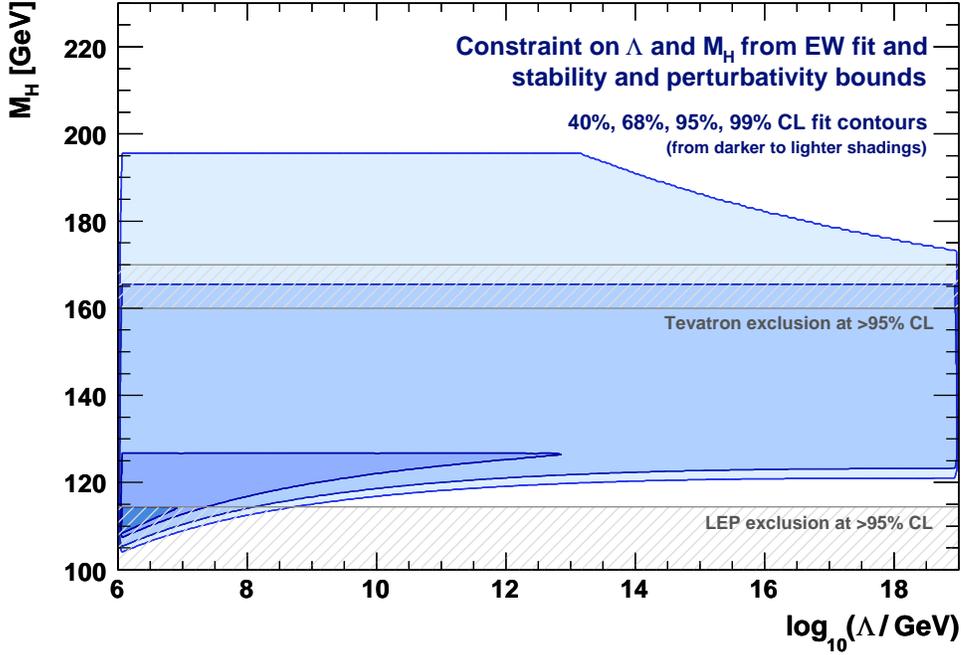, scale=\defaultFigureScale}\vspace{0.5cm}

  \epsfig{file=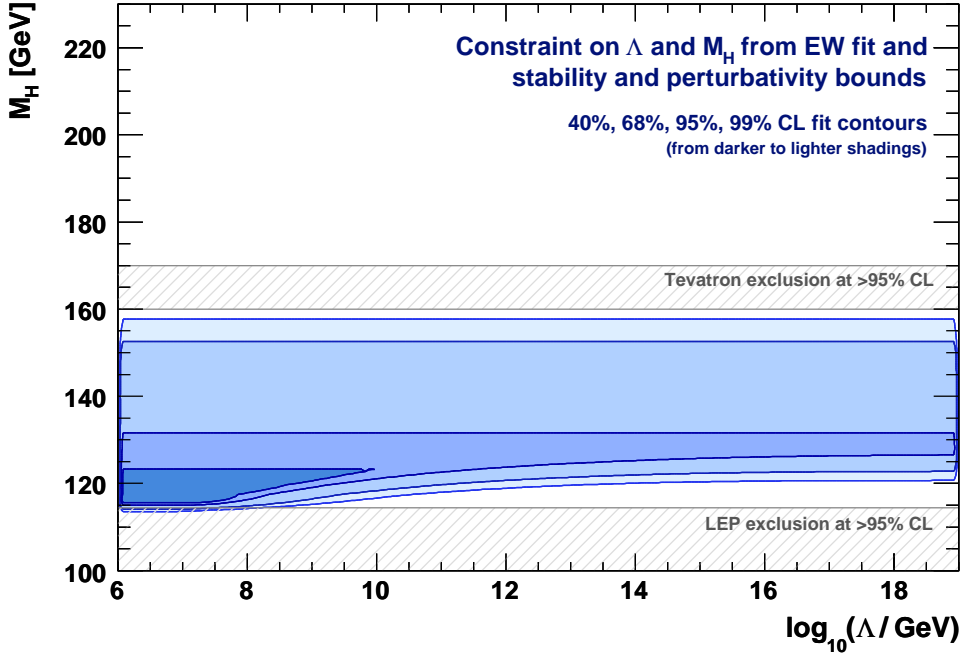, scale=\defaultFigureScale}
\end{center}
  \vspace{-0.3cm}
\caption{\it Contours of 40\%, 68\%, 95\% and 99\%~CL obtained from scans 
            of fits with fixed values of the variables $M_H$ and 
            $\log_{10}(\Lambda/\gev)$. The fits include the electroweak 
            precision data and the bounds from the perturbativity and 
            stability requirements shown in Fig.~\protect\ref{fig:bounds}. 
            The lower plot also incorporates the direct Higgs boson searches
            at LEP and the Tevatron (corresponding to the complete fit scenario
            in Ref.~\cite{gfitter}). Their respective 95\%~CL exclusion domains 
            are depicted by the hatched bands.}
\label{fig:mh-vs-lam}
\end{figure}

The plots in Fig.~\ref{fig:mh-vs-lam} show the constraints obtained in the
two-dimensional plane $M_H$ versus $\log_{10}(\Lambda/\gev)$ from combined
fits excluding (upper plot) and including (lower plot) the direct Higgs
searches, respectively. The shaded bands indicate the 40\% (innermost,
darkest), 68\%, 95\% and 99\% (outermost, lightest)  CL allowed regions.\footnote
{
   Although the test statistic in Fig.~\ref{fig:mh-vs-lam} corresponds in 
   principle to two degrees of freedom, an effective constraint on $\log_{10}(\Lambda/\gev)$ 
   only occurs along the bounds, so the number of degrees of freedom in the majority of 
   the plane is one. This is the value we have used to translate the test
   statistics into the $1-{\rm CL}$ values via ${\rm Prob}(\Delta\chi^2,n_{\rm dof})$. 
   A complete analysis would require the generation of very large numbers of toy Monte Carlo 
   measurements, which is beyond the scope of this paper. (Such a study has been 
   performed in Ref.~\cite{gfitter} in the framework of a Two-Higgs-Double Model 
   analysis.) 
}
We find that the overall $\chi^2$ estimator has the following minimum
values in the planes depicted: 17.2 (excluding the direct Higgs searches) 
and 17.8 (including the direct searches).\footnote
{
   The difference in the former number with respect to Ref.~\cite{gfitter}
   (16.4) is due to the restriction to $M_H>100\gev$ and $\Lambda>10^6\gev$ 
   imposed here.
}
The overall fit is of satisfactory quality for the 13 (14) degrees of
freedom excluding (including) the direct Higgs constraint, and we see no
need to doubt that the SM is a suitable framework for analysing the
available electroweak data (\cf the statistical analysis and discussion 
in Sec. 4.2.3 of Ref.~\cite{gfitter}).

The values of $M_H$ favoured by the global fit are compatible with a value
of the SM cut-off scale $\Lambda$ up to the Planck scale. Only for Higgs
masses below $124\gev$ or above $172\gev$ would the bounds provide a
constraint on $\Lambda$. Because of the small dependence of the stability 
bound for $M_H$ on $\Lambda$, its theoretical uncertainty significantly 
impacts the value of the constraint obtained.

The Tevatron results do however increase our confidence
that, within the SM, the Higgs quartic coupling is perturbative up to
$M_P$.  Without the direct Higgs searches, the $\Delta\chi^2$ price is
$4.1$ for $M_H$ falling into the `blow-up' region, which -- assuming a
proper $\chi^2$ behaviour -- translates into an exclusion of the `blow-up'
region at the 95.7\%~CL.  Including the Tevatron Higgs results leads to a
higher $\Delta\chi^2$ price of $6.9$, corresponding to an improved
exclusion at the 99.1\%~CL.  {\it Hence the SM probably does not blow up 
before the Planck scale.}

\begin{figure}[t]
\begin{center}
  \epsfig{file=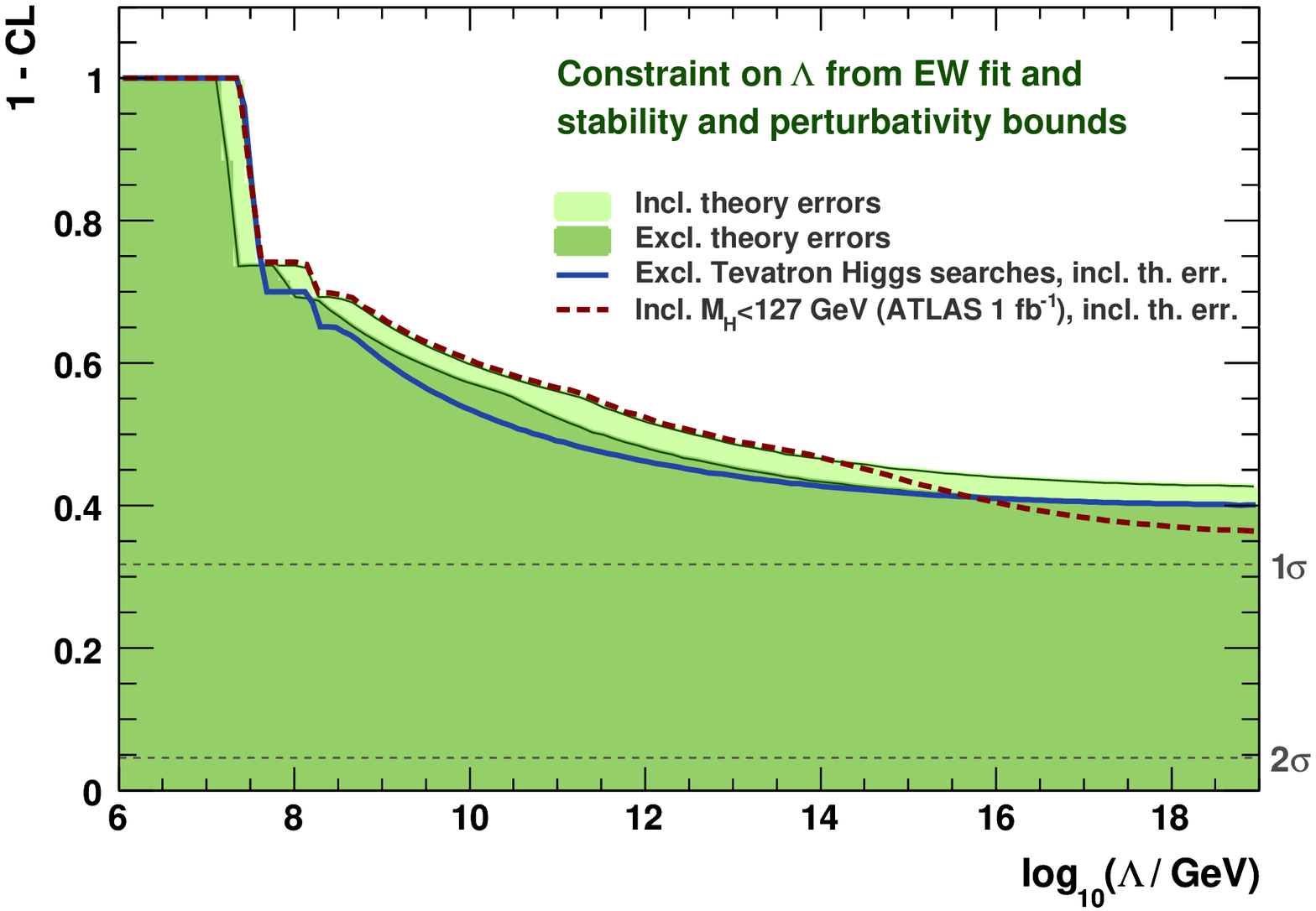, scale=\defaultFigureScale}
\end{center}
  \vspace{-0.5cm}
\caption{\it Constraint on $\Lambda$ from the global electroweak fit and the requirement
         of absolute vacuum stability and perturbativity, expressed as 1\,$-$\,CL
         and assuming it to be given by Prob($\Delta\chi^2$,1). Shown are fits
         with (light shading) and without (dark shading) taking into account 
         the theoretical uncertainty in the stability bound. The bold solid [blue] 
         line shows the effect of removing the Tevatron Higgs searches from the 
         global fit. The dashed [red] line shows the effect of a 
         hypothetical upper bound of $M_H < 127\gev$ at 95\%~CL, as might 
         be obtained with early data at the LHC.}
\label{fig:lam}
\end{figure}

The result of the global fit as a function of $\Lambda$ can be used to
assess the p-value of the `survival' scenario. Figure~\ref{fig:lam}
shows $1-\CL$ versus $\Lambda$ for various cases: with and without the
theoretical uncertainty in the stability bound, including and excluding
the Tevatron Higgs results, and assuming a hypothetical unsuccessful early
Higgs search at one of the high-$p_T$ LHC experiments (represented here by
ATLAS), for an integrated luminosity of approximately $1\:{\rm fb}^{-1}$
at 14\tev centre-of-mass energy, that should have sufficient sensitivity
to exclude $M_H > 127\gev$ at 95\%~CL~\cite{atlas-cscbook}.

\begin{figure}[t]
\begin{center}
  \epsfig{file=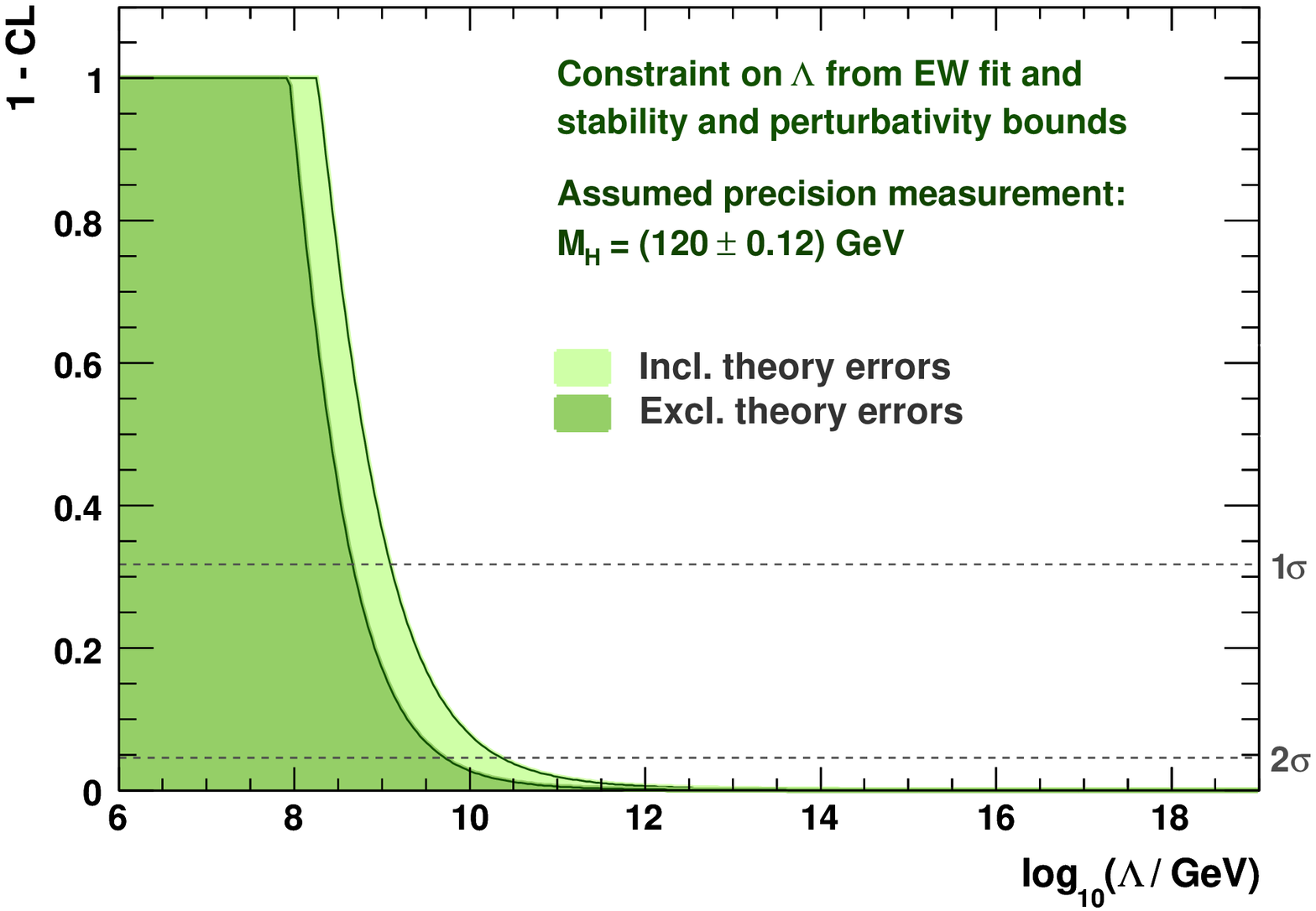, scale=0.405}
  \epsfig{file=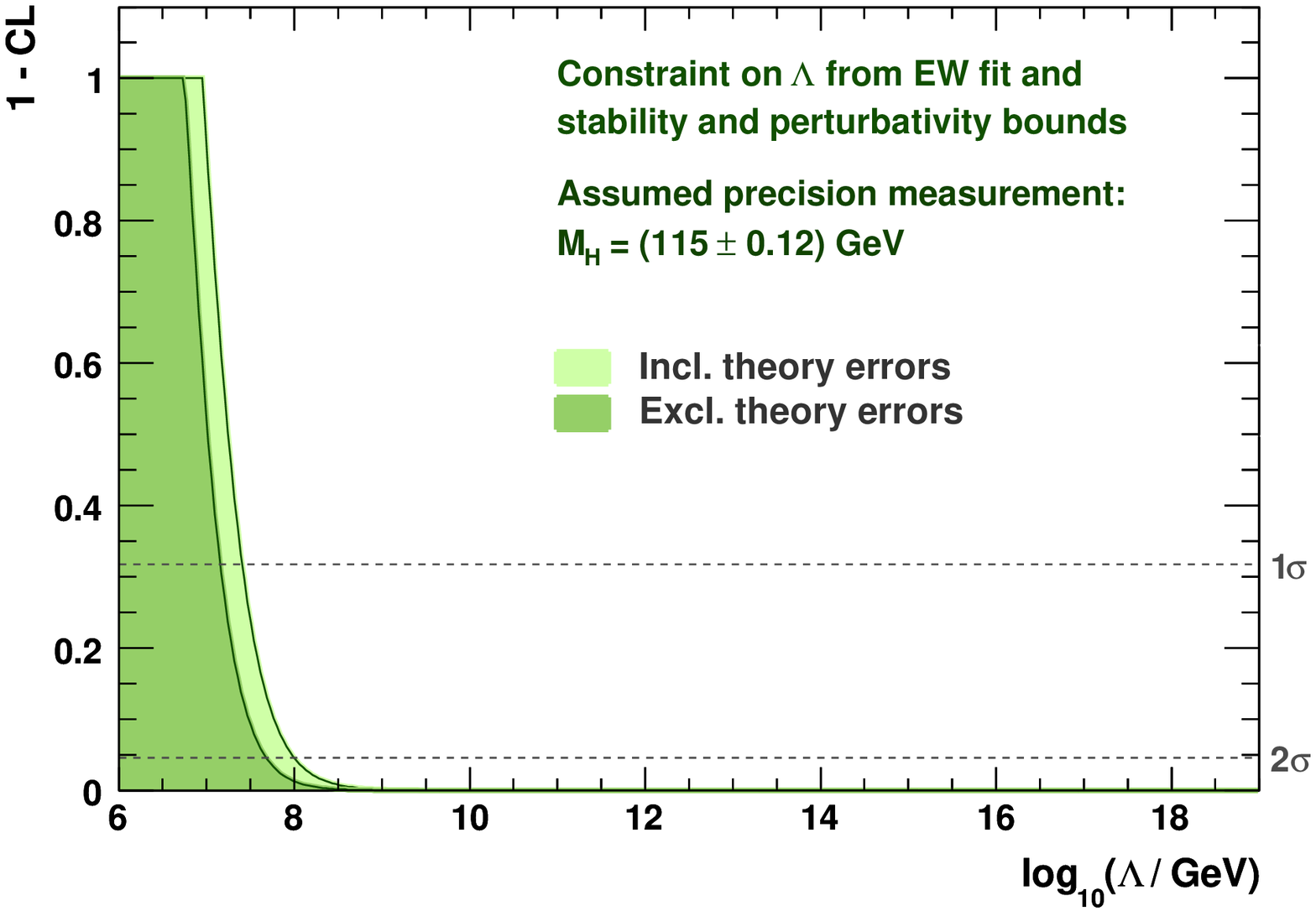, scale=0.405}
\end{center}
  \vspace{-0.5cm}
\caption{\it Constraint on $\Lambda$ from the global electroweak fit and the 
         requirement of absolute vacuum stability and perturbativity. Included 
         in the fit is a hypothetical Higgs discovery
         with precise (0.1\%) mass measurement at $M_H=120\gev$ (left plot) and 
         $M_H=115\gev$ (right plot), respectively. Shown are fits
         with (light shading) and without (dark shading) taking into account 
         the theoretical uncertainty in the stability bound. 
         Also included are improved errors for the top and $W$ masses, as
         anticipated for the LHC (see text).}
\label{fig:lam_mh}
\end{figure}

No constraint on $\Lambda$ (assuming absolute stability) that would reach or 
exceed 68\%~CL can be derived from
the present data, nor from the prospective incremental improvement in the
Higgs constraint that might come from the Tevatron or the early
running of the LHC. If, however, there were a Higgs
discovery with a mass determined to be $M_H=120\gev$ 
or $M_H=115\gev$ (assumed precision 0.1\%) after years of successful LHC 
operation, one would obtain the constraints on $\Lambda$ plotted in
Fig.~\ref{fig:lam_mh}. For these plots, we have also included prospectives 
for the precision of the top and $W$ mass measurements of $1\gev$ and 
$15\mev$ overall errors, respectively (see references in~\cite{gfitter}). 
The 95\%~CL upper limits on the cut-off scale obtained including theoretical errors would read
$\log_{10}(\Lambda/\gev)<10.4$ and $8.0$, respectively, including an almost half an 
order of magnitude theoretical uncertainty. In this case, one would obtain an
upper limit on the absolute stability of the SM that would be comparable
with the scale suggested by the seesaw model for the light neutrino
masses. The p-values of the $M_H=120$ and $115$\gev scenarios for the 
`survival' up to $M_P$ are as small as the occurrence of $3.5\sigma$ and 
$5.3\sigma$ fluctuations, respectively.  

\section{Conclusions}

We have combined a global fit of the available electroweak data to the SM
and results from direct searches for the SM Higgs boson
with theoretical calculations of the effective Higgs potential, using
two-loop RGEs to extrapolate its behaviour to high scales. Our analysis
displays the impact of the most recent Tevatron searches for an
intermediate-mass Higgs boson. We find an exclusion at the 99.1\%~CL
of the possibility that the quartic Higgs coupling of the SM could blow up
at some scale $\Lambda$ below the Planck scale, which the Tevatron data
have increased from the 95.7\%~CL found with the precision electroweak
data alone.

On the other hand, the present data exhibit no clear preference between
scenarios in which the SM survives up to the Planck scale, and in which it
develops new minima at a scale $\Lambda$ and becomes metastable with
respect to either thermal or zero-temperature fluctuations. Here the Tevatron data do not
change greatly the {\it status quo ante} even though they reduce the `survival' region.
Nor would a hypothetical LHC
upper limit $m_H < 127\gev$ nor, {\it a fortiori}, hypothetical
incremental improvements in the Tevatron upper limit on Higgs production.
However, discovery of the Higgs boson might reveal quite conclusively the
possible fate of the SM. For example, if the SM Higgs boson were to be
discovered with a mass of 120 (115)~GeV, the effective potential of the SM
would develop a new vacuum at $\log_{10}(\Lambda/\gev)<10.4 (8.0)$ and
remain in a metastable state, unless new physics beyond the SM intervenes.
Needless to say, our considerations might be happily irrelevant  if LHC 
finds direct evidence for new physics at some scale $\Lambda$.

\section*{Acknowledgements}
We are indebted to the Gfitter Group\footnote
{
  M.~Baak, H.~Fl\"acher, M.~Goebel, J.~Haller, A.H., D.~Ludwig, 
  K.~M\"onig, M.~Schott, J.~Stelzer.   
} 
for making the Gfitter package available, and for the vast collection of 
experimental results used for the numerical analysis presented in this 
paper. We also thank Gino Isidori for a useful discussion on metastability 
bounds. J.R.E. thanks CERN for
hospitality and partial financial support. His work is supported in part 
by CICYT, Spain, under contracts FPA2007-60252; by a Comunidad de Madrid 
project (P-ESP-00346); and by the European Commission under contracts 
MRTN-CT-2004-503369 and MRTN-CT-2006-035863.

\end{document}